\documentclass[10pt,twocolumn]{article}
\usepackage{graphicx}
\usepackage{subfigure}
\usepackage{relsize}
\usepackage[a4paper,margin=2cm]{geometry}
\usepackage{amssymb}
\usepackage{xfrac}
\usepackage{ifthen}
\usepackage{amsmath}
\usepackage{amsfonts}
\usepackage{upgreek}
\usepackage[version=3]{mhchem}
\usepackage{bm}% bold math
\usepackage{multirow}
\usepackage{color}
\usepackage{placeins}
\usepackage{todonotes}

\usepackage{tikz}
\usetikzlibrary{arrows}
\usetikzlibrary{patterns}
\usetikzlibrary{external}
\begin{document}

\title{Using LHC Forward Proton Detectors to Tag Nuclear Debris}
\author{
Rafa\l{} Staszewski\thanks{Corresponding author (rafal.staszewski@ifj.edu.pl)}\ \   and Janusz J. Chwastowski\\[7mm]
Institute of Nuclear Physics Polish Academy of Sciences\\
ul. Radzikowskiego 152, 31-342 Krak\'ow, Poland\\[7mm]
}

\maketitle
\begin{abstract}
  The forward proton detectors, already existing at the LHC, are considered in the context of heavy ion collisions.
  It is shown that such detectors have the potential to measure nuclear debris originating from spectator nucleons.
  The geometric acceptance for different nuclei is studied, and how it is affected by the motion of the nucleons in the nucleus and by the experimental conditions.
  A possibility of extending the acceptance region is discussed.

\end{abstract}

\section{Introduction}

The Large Hadron Collider \cite{bib:lhc} is equipped with dedicated detectors allowing measurements of protons scattered in diffractive or electromagnetic interactions.
Since the scattering angles of such protons are very small, these detectors are installed very far away from the interaction point (IP).
% -- typically around 200 m.
In addition, due to the use of the roman pot technology, they can be placed very close to the proton beams.

The LHC physics program is not solely devoted to the studies of the proton--proton interactions.
Possibilities of accelerating heavy ion beams have already resulted in many measurements of proton--lead and lead--lead collisions \cite{bib:qm2014}.

\begin{figure}[thb]
  \centering
  \includegraphics{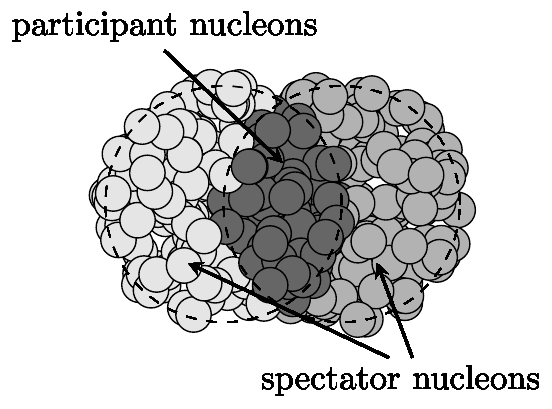}
  \caption{A schematic diagram of a ultra-relativistic heavy ion collision (view perpendicular to the relative velocities).}
  \label{fig:AA_interaction}
\end{figure}

An ultra-relativistic interaction of two heavy nuclei is sketched in Fig. \ref{fig:AA_interaction}.
Typically, the impact parameter has a non-zero value and only a part of the nucleons constituting one nucleus collides with a part the nucleons 
of the other nucleus.
The nucleons actively participating in the interaction are called the participant nucleons (or participants),
in contrast to the spectator nucleons (spectators).

Since the time-scale of the ultra-relativistic ion collision is much shorter than that of the interactions within the nucleus,
the spectators are essentially left intact during the nucleus--nucleus collisions.
They are scattered into the accelerator beam pipe and escape the acceptance of the detectors, very much like diffractive protons.

%In this work we study if
The present work tries to answer whether and to what extent the forward proton detectors at the LHC could be used with heavy ion beams.
The paper is organised as follows. In Section \ref{sec:detectors} the forward proton detectors at the LHC are introduced.
Section \ref{sec:methodology} describes the methodology of the presented study.
The results are presented in Section \ref{sec:results} and  the conclusions in Section \ref{sec:conclusions}.

\FloatBarrier
\section{AFP Detectors}
\label{sec:detectors}

\begin{figure*}[t]
  \centering
  \includegraphics[width=0.9\linewidth]{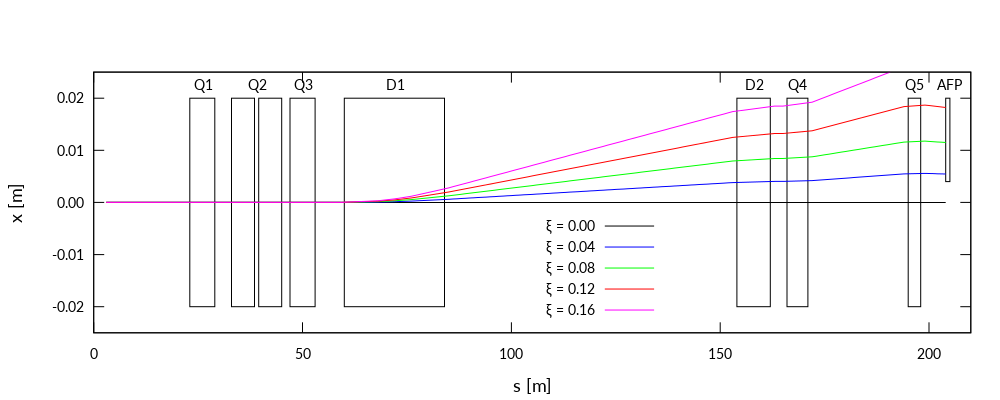}
  \caption{ 
  Trajectories of diffractively scattered proton in the LHC magnetic fields. 
  $x$ is the horizontal coordinate of the trajectory with respect to the nominal orbit,
  $s$ is the distance from the interaction point along the orbit,
  $\xi$ is the relative energy loss of the protons.% From \cite{CieslaStaszewski}.
  }
  \label{fig:trajectories}
\end{figure*}

Several systems of forward proton detectors are installed at the LHC. These are: TOTEM~\cite{bib:totem}, CT-PPS~\cite{bib:pps}, ALFA~\cite{bib:alfa} 
and AFP~\cite{bib:afp}.  All these detectors are placed about 200 m away from their corresponding interaction points.
The ALFA detectors approach the beams in the vertical direction, the AFP and CT-PPS approach horizontally, while TOTEM has both types of detectors.
The present analysis focusses on the AFP detectors. However, the results for other horizontal ones can be expected qualitatively similar.
The potential application of the vertical detectors is not to be covered within this work.

The AFP (ATLAS Forward Proton) detectors \cite{bib:afp} are a sub-system of the ATLAS experiment set-up located at the LHC Interaction Point 1. 
It is foreseen to consist of four stations -- two on each outgoing beam. The near stations will be placed 205 m and the far ones 217~m away from the 
interaction point.
One arm of the AFP detector system, the stations on one side of the IP1, has already been installed. The installation of the other arm is planned for the 
2016/2017 LHC technical stop.

Each AFP station consists of a roman pot mechanism allowing for the horizontal insertion of the detectors into the accelerator beam pipe.
Each station contains a tracking detector of four planes of the 3D Silicon sensors \cite{bib:Si}. 
Additionally, the far stations will be equipped with quartz-based Cherenkov time-of-flight counters.
However, these counters are not relevant for the present study.

A diffractively scattered proton, before being detected in AFP, passes through the magnetic fields of seven LHC magnets.
The Q1 -- Q3 triplet of quadrupole magnets is responsible for the final focusing and the emittance matching of the beams, providing thus the high 
luminosity.
The two consecutive magnets are the dipoles: D1 and D2. They separate the incoming and the outgoing beams and accommodate them within the 
corresponding beam pipes of the machine. The last two quadrupole magnets, Q5 and Q6, are used to match the beam optics in the interaction region to 
the rest of the ring.

The momentum of the diffractively scattered proton slightly differs from that of the beam proton. 
In the interaction the proton looses a part of its energy and its transverse momentum is, on the average, increased.
It should be recalled that the distribution of the transverse momentum of diffractively scattered protons is very steep.
Therefore, it is the scattered proton energy which mainly determines its trajectory and the position in the detector.
The transverse momentum of a typical magnitude only leads to a moderate smearing of this position. 

%Also, its transverse momentum is, on the average, 
%greater, because it has been scattered. In addition, its energy is smaller, because it was lost in the interaction to produce additional particles.

Lower energy of the diffractive proton means that the curvature of its trajectory in the magnetic field will be greater, which will cause the scattered proton to recede from the beam orbit. 
This property allows the measurement of such protons with detectors placed close to the beam. 
As an example, Fig. \ref{fig:trajectories} shows the trajectories of protons with different energy values, here specified by the relative energy loss $\xi = 1-E_{proton}/E_{beam}$.
Also, the LHC magnets and the AFP detector are depicted in this figure.

The kinematic range in which the measurements are possible can be quantified by the value of the geometric acceptance as a function of the proton 
energy and its transverse momentum (here the acceptance is averaged over the azimuthal angle). 
Naturally, the acceptance depends on the exact settings of the magnetic fields, the so-called machine optics, and on the beam--detector distance.

\begin{figure}[htbp]
  \centering
  \includegraphics[width=\linewidth]{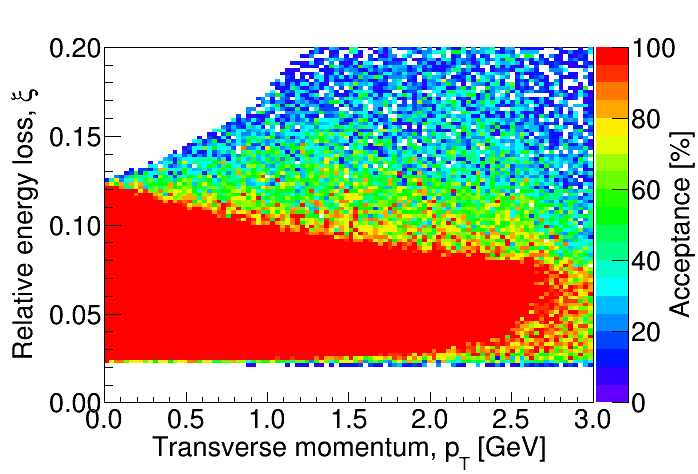}
  \caption{
  Geometric acceptance of the AFP detectors for diffractive protons.
  %$\xi$ is the fraction of energy lost by the proton.
  %From \cite{CieslaStaszewski}.
  }
  \label{fig:proton_acceptance}
\end{figure}

Fig. \ref{fig:proton_acceptance} presents an exemplary acceptance plot, calculated for 7~TeV proton beams and the $\beta^\ast = 0.55$~m LHC optics 
\cite{bib:lhcoptics}. The presented results were obtained %\cite{CieslaStaszewski}
with the Mad-X programme \cite{bib:madx}.
One can see that the AFP detectors allow an efficient measurement of protons that lost between 3\% and 12\% of its energy and gained less than 
2.5~GeV of transverse momentum. 

%As will be discussed in the following, a similar situation takes place when the transport of nuclear debris is considered.

\FloatBarrier
\section{Methodology}
\label{sec:methodology}

The spectator nucleons after an interaction of ultra-relativistic heavy ions are left in a very peculiar state.
Before the collision the nucleons were a part of the nucleus, interacting with other nucleons belonging to it.
Then, the participating nucleons are ``taken away'' and the spectators are left as members of a highly excited ensemble,
which subsequently decays into lighter fragments.

The discussion concerning the details of this process is outside the scope of the present study, see for example \cite{ferrari, dpmjet, olson, frankland,jacak} and references therein.
In the following it is assumed that all the nuclei lighter than the projectiles can possibly be produced without paying any attention to the abundances of 
particular isotopes. % created in the high energy ion-ion collision.
Next, it is studied which of the produced nuclei could be detected by the AFP system.

%such nuclei (totally ionisated) were tracked through the LHC magnetic lattice to find 
%out which could be detected by the AFP system.
%Then, we find out which of them could be detected.

\begin{figure}[htbp]
  \centering
  \includegraphics[page=2, width=\linewidth]{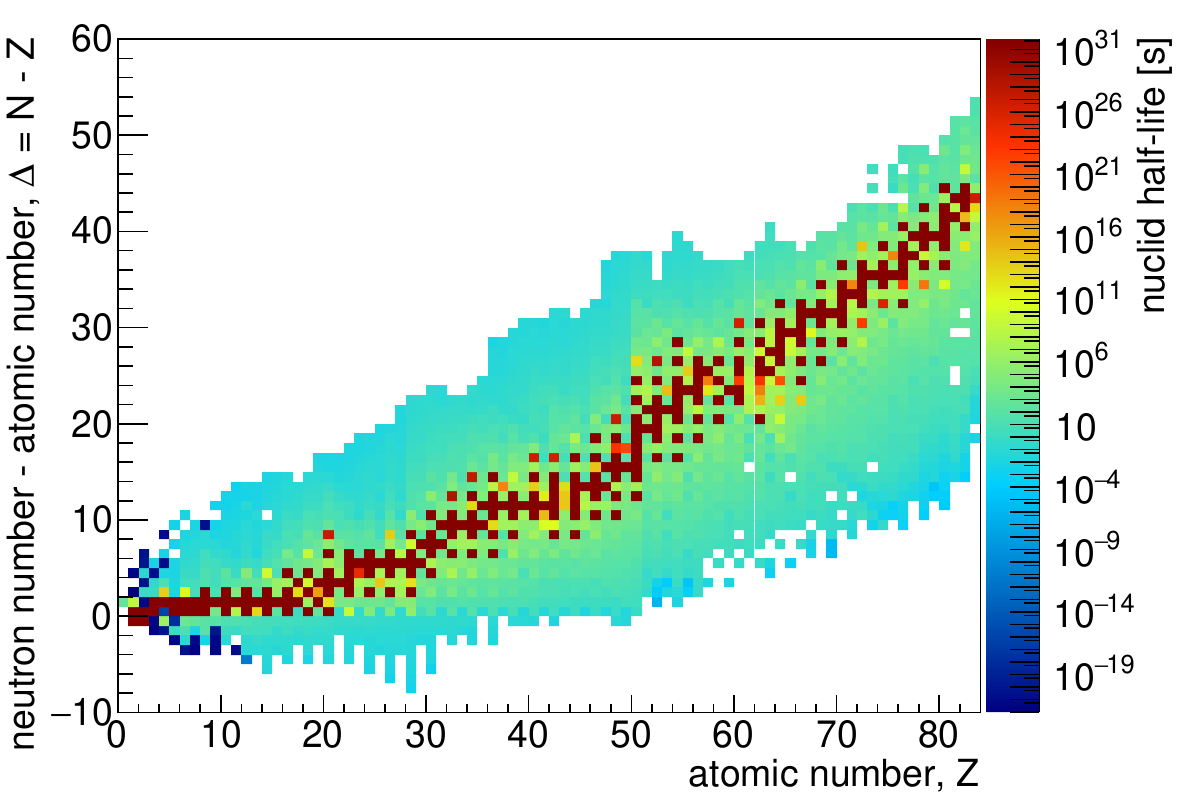}
  \caption{Half-life of known nuclei.}
  \label{fig:decay}
\end{figure}

Since the detectors are positioned quite far away from the interaction region, it is worth checking which produced nuclei have a chance to reach the AFP stations before they decay.
% from the interaction point to the  detector located at the distance of about 200 metres. 
The spectators move with velocity close to the speed of light
and their Lorentz factor is $\gamma=2751$  (for the lead beam and the LHC magnets set as for 6500 GeV proton beams).
%
% 208Pb mass = 207.98 u
% u = 931.494 MeV
% m = 193.732 GeV
% for proton energy og 6500 GeV Pb will be accelerated to 82*6500 = 533000 GeV
% gamma = E/m = 2751
Then, the proper time of a nucleus before it reaches the AFP detector is less than a nanosecond.
Fig. \ref{fig:decay} presents the half-life times of the known nuclei%. A reader should note that 
%\footnote{%In this paper various variables are plotted in a non-standard way,
\footnote{This and other plots in this paper are presented in a non-standard way, as a function of the atomic number $Z$ and the difference of the number of neutrons, N, and the atomic 
number, $\Delta = N - Z = A - 2Z$, where $A$ is the mass number.}.
%This choice of variables makes a better use of the plot area.
%}.
It is clear that the vast majority of the nuclei could reach the detectors.

If one neglects the internal motion of nucleons within the nucleus, then all the nucleons carry the same energy equal to the energy of the beam divided
by the mass number of the beam particles: $E_N = E_b/A_b$. Assuming that the spectators are left intact in the collision,
a nuclear fragment with mass number $A$ consisting of the spectators would have the energy equal to $A\cdot E_N$.

In order to take into account the internal motion of nucleons, the Fermi-gas model of a nucleus was employed.
In the rest frame of the nucleus, the density of nucleon states is given by $\text{d} n \sim p^2 \text{d} p$.
Therefore, in the simulation the absolute value of the momentum of each nucleon was randomly drawn from a quadratic distribution between zero and the Fermi momentum of 250~MeV.
The direction of each momentum was assumed to be isotropic. The momentum of a given fragment was calculated as a vector sum of 
momenta of all its nucleons. Finally, the nucleus momentum was Lorentz-transformed into the laboratory frame.

It is worth to emphasize the main points of the presented methodology. First, no physical model of the nuclear debris production is used.
This is a clear drawback, because the results will not say whether the AFP detectors are expected to register anything.
However, such models would require extrapolations of the low energy measurements and it is not certain that such an extrapolation would provide
reliable results.
On the other hand, with the approach chosen here, the results are essentially model independent and can be applied on top of any particular model.
The only exception are the small details of the Fermi motion, where possible collective movements inside the nuclei are neglected.
However, this can only have a small effect on the presented results.

In order to simulate the transport of the nuclei through the accelerator structures%
\footnote{The LHC optics files are available \cite{bib:lhcoptics} in the Mad-X format, 
while, up to the knowledge of the authors,
Mad-X does not allow simulating trajectories of particles of type different than that of the beam.}
a simple observation was employed: the trajectory in the magnetic field depends on the ratio of the particle momentum to its charge.
Therefore, the trajectory of a particle with momentum $\mathbf{p}$ and charge $q$ is the same as the trajectory of the beam particle with 
charge $q_b$ and momentum:
\begin{equation*}
  \mathbf{p'} = \frac{q_b}{q}\,\mathbf{p},
  %\label{eq:momentum}
\end{equation*}
assuming $q$ and $q_b$ have the same sign, which is true for all nuclei. 

Results presented below were obtained assuming that the beam of fully ionised \ce{^{208}_{82}$Pb$} ions is accelerated to the energy of 2.56 TeV per nucleon, which corresponds to 6.5 TeV energy protons.
During the LHC RUN 2, the data-taking involving the heavy ion beams is performed with the $\beta^\ast = 0.8$~m optics~\cite{bib:lhcoptics}.

\section{Results}
\label{sec:results}

Neglecting the spreads due to the beam emittance and the Fermi motion, a nucleus with a given $A$ and $Z$ numbers will hit the AFP detectors in a 
well defined position.
Since the AFP detectors approach the beam in the horizontal direction, it is the $x$ coordinate of the nuclear fragment trajectory which plays the major 
role in the present considerations. One should recall that for the safety reasons the detector is positioned at some distance with respect to the 
beam and that there is additional dead material of the roman pot floor.
%of the nucleus position that defines whether it can or cannot be defined.

\begin{figure}[htbp]
  \centering
  \includegraphics[page=4, width=\linewidth]{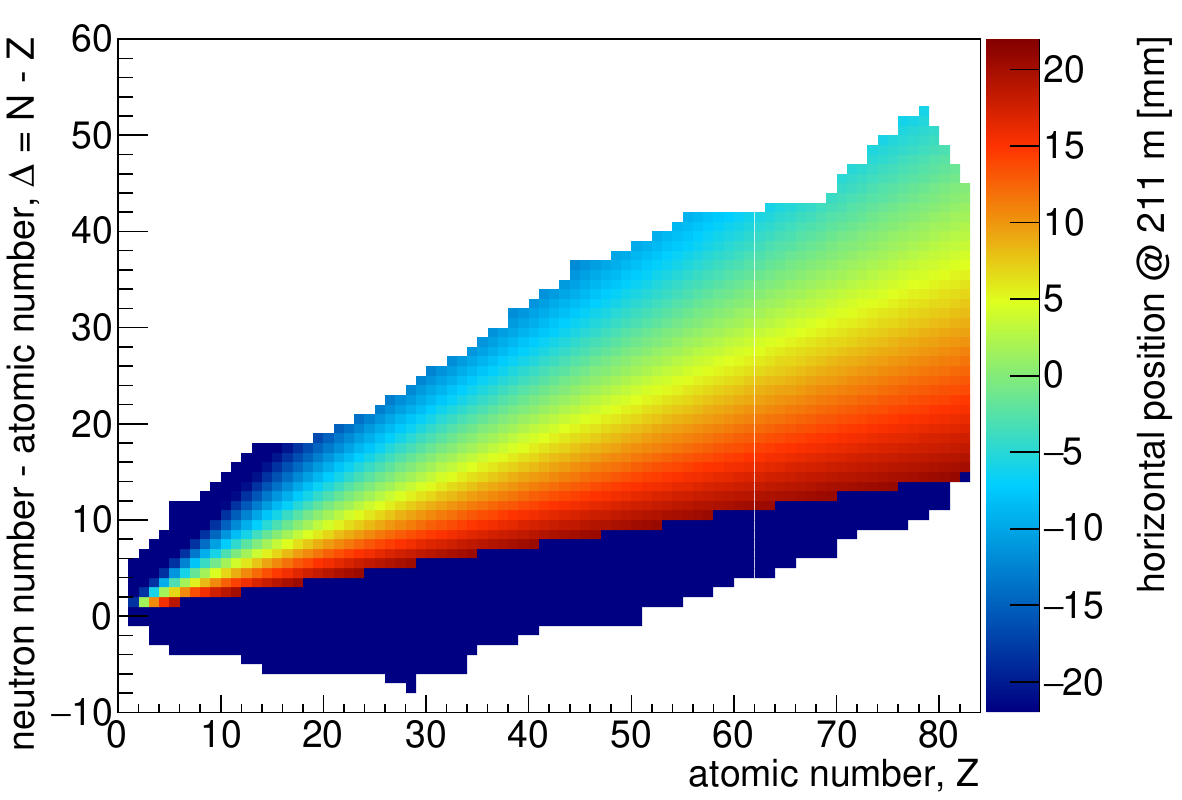}
  \caption{
  Horizontal position of nuclei at 211 m from the ATLAS interaction point.
  The nuclei lost in the LHC apertures are drawn in dark blue.
  }
  \label{fig:positions}
\end{figure}

Fig. \ref{fig:positions} shows the horizontal positions of the nuclei at 211 m from the interaction point (in the middle between the two AFP 
stations) within the accelerator beam pipe.
The position of \ce{^{208}_{82}$Pb$} ($\Delta=44$) and all nuclei with the same $A/Z$ ratio (\textit{i.e.} the same $\Delta/Z$ ratio) is equal to zero.
 %, as should be expected from the beam properties.
Nuclei with less neutrons per proton  are deflected outside the LHC ring, similarly to diffractive protons, and can be registered the AFP 
detectors. Nuclei containing more neutrons per proton are deflected towards the LHC centre escaping the detection.
Nuclei with $A/Z$ very different from lead can be lost in the LHC apertures and do not reach 211 m at all.

For the %\textbf{??normalised??} 
emittance value of 1.233~$\upmu$m \cite{bib:lhcoptics} the lead beams at the interaction point have angular spread of 24~$\upmu$rad,
% sqrt(1.233e-6/0.8/2751)
while the interaction vertex distribution has the transverse spread of 13~$\upmu$m
and the longitudinal one of 5.5~cm.
% sqrt(1.233e-6*0.8/2751)/sqrt(2)
The horizontal width of the beam at 211~m from the interaction point, $\sigma_x$, is equal to 134~$\upmu$m.
% beta ~= 40 m, sqrt(1.233e-6*40/2751) 
This width is a usual unit of the distance between the detector and the beam. % following from the accelerator and the detector safety considerations.  
%For safety reasons, the AFP detectors cannot operate too closely to the beam.
In the following, it was assumed that the sensitive area of the sensor is placed 3~mm from the centre of the beam.
This distance covers about $19\sigma_x$ and 
%which is approximately equal to $22\sigma_x$ and it 
a 0.5 mm-long distance between the active detector edge and outer side of the roman pot floor.

\begin{figure}[t]
  \centering
  \includegraphics[page=5, width=\linewidth]{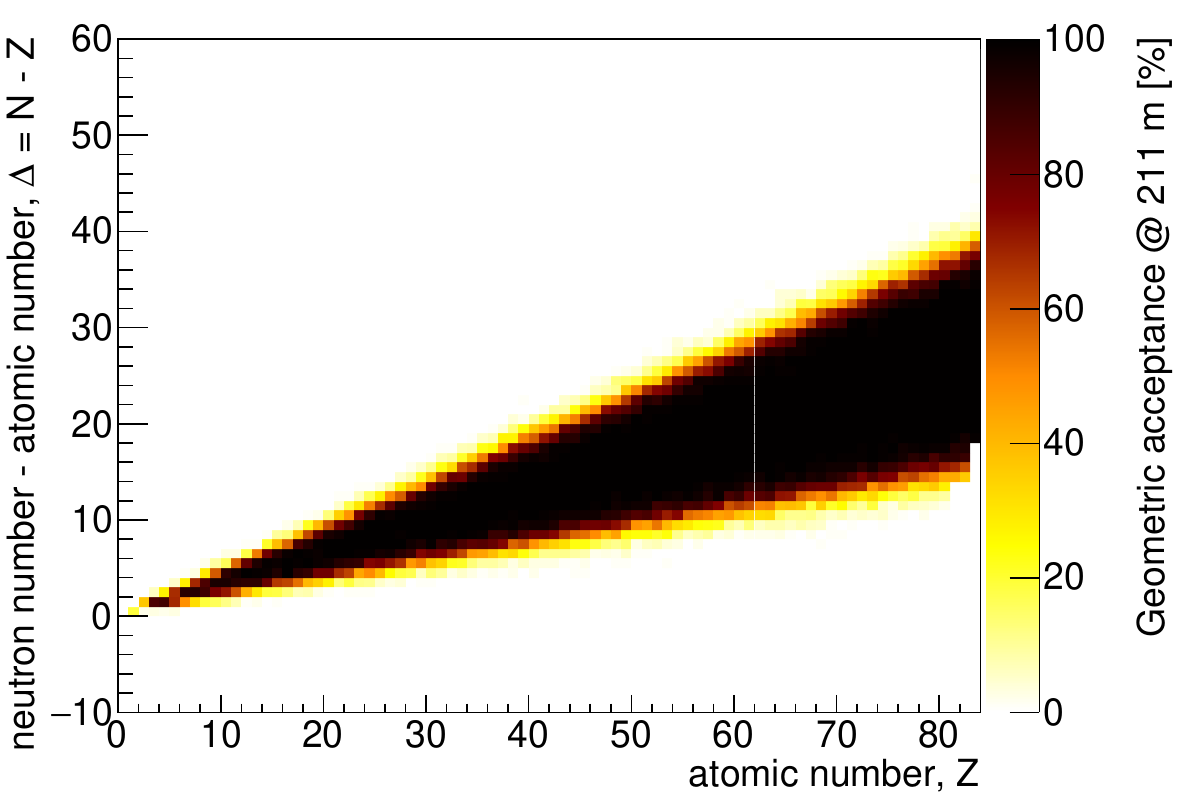}
  \caption{AFP acceptance for nuclear debris.}
  \label{fig:acceptance}
\end{figure}

Fig. \ref{fig:acceptance} shows the acceptance to detect a given nucleus as a function of $(Z, \Delta)$.
The obtained results were averaged over the distributions of momenta discussed before and the Gaussian spreads of the LHC beams (spatial and angular).
The observed shape is in agreement with what one would expect from Fig. \ref{fig:positions}.

Although, the AFP detectors were not designed and build to detect the nuclear debris,  their acceptance covers a significant part of the nuclei spectrum.
This is particularly true for the heavier nuclei, where for a given $Z$ more than a half of known nuclei can be potentially detected.
With decreasing $Z$ value the range of the accepted masses  linearly decreases.

\begin{figure}[t]
  \centering
  \includegraphics[width=\linewidth]{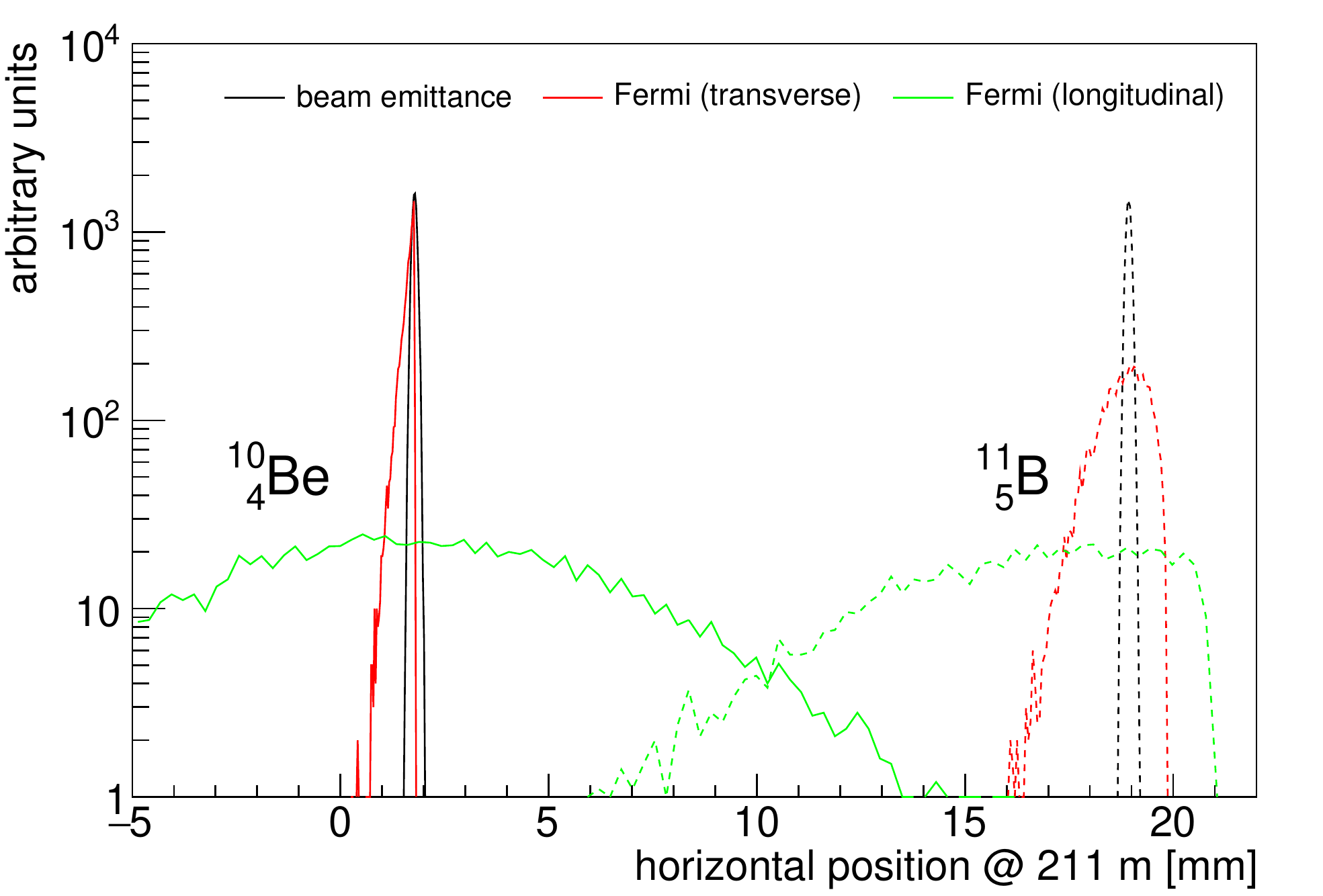}\par
  \includegraphics[width=\linewidth]{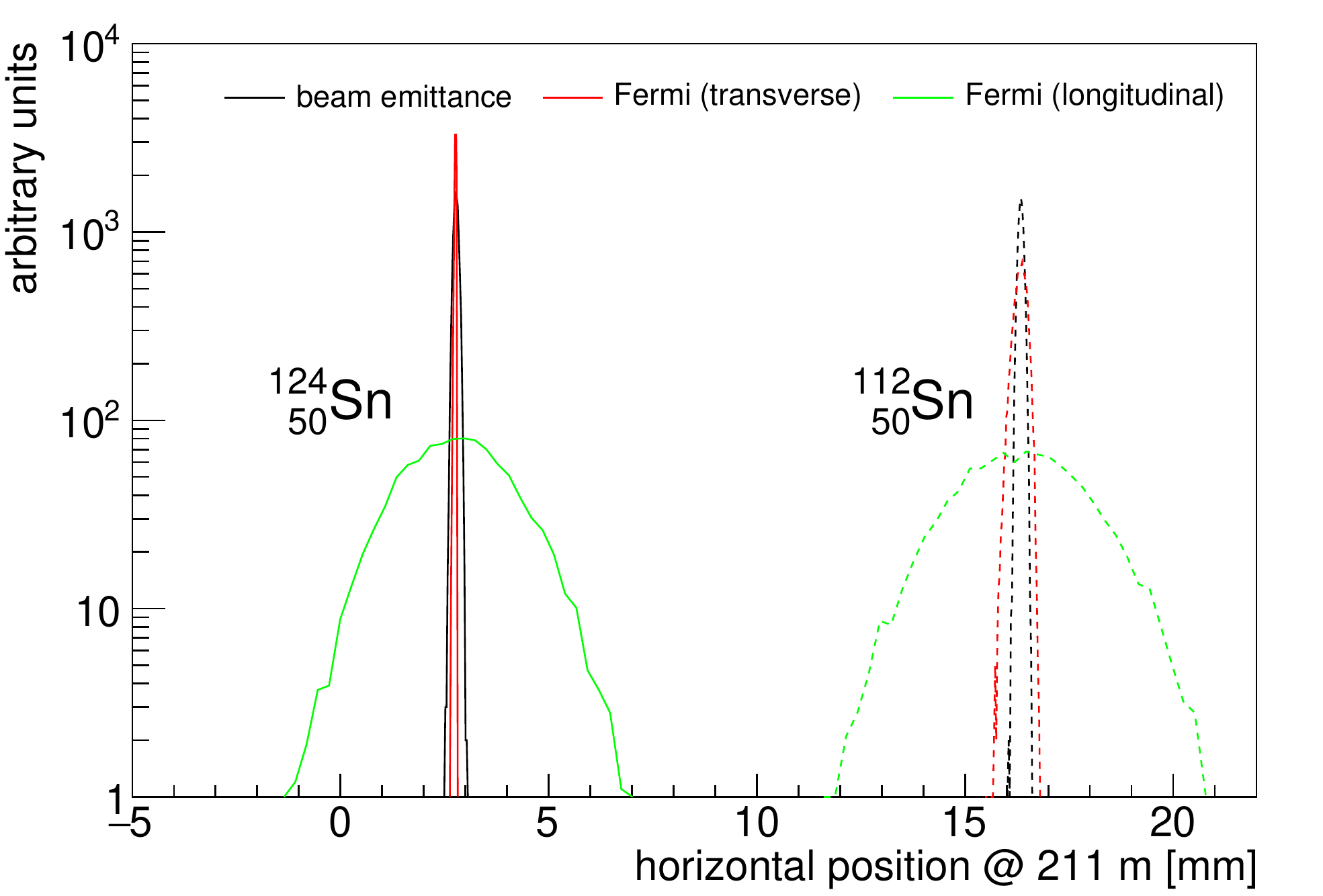}
  \caption{Effects of finite beam emittance and the Fermi motion on the fragment position in AFP for exemplary light (top) and heavy (bottom) nuclei.}
  \label{fig:effects}
\end{figure}

A study of the influence of various spreads on the position of selected ions at the distance of 211 m away from the IP is presented in Fig.~\ref{fig:effects}.
As can be observed the effects of the beam spreads and those due to the transverse component of the Fermi motion are rather small, but clearly 
visible.
The ion position smearing is dominated by the longitudinal Fermi motion magnified by the Lorentz boost. One can observe that this effect is stronger for 
lighter nuclear debris. 
%This means that the influence of the spreads is rather small, but clearly visible, and the behaviour is predominantly governed by the Lorentz boost.

A natural question arises whether it is possible to improve or modify the AFP acceptance to measure the spectator fragments.
One key parameter is the distance between the detectors and the centre of the beam.
With detectors operating at a smaller distance, the acceptance for nuclei with higher $A/Z$ ratio would be increased.
Another option would be a modification of the machine optics, which may be considered in the future, if the measurements proposed in this paper prove useful.

\begin{figure}[t]
  \centering
  \includegraphics[width=\linewidth]{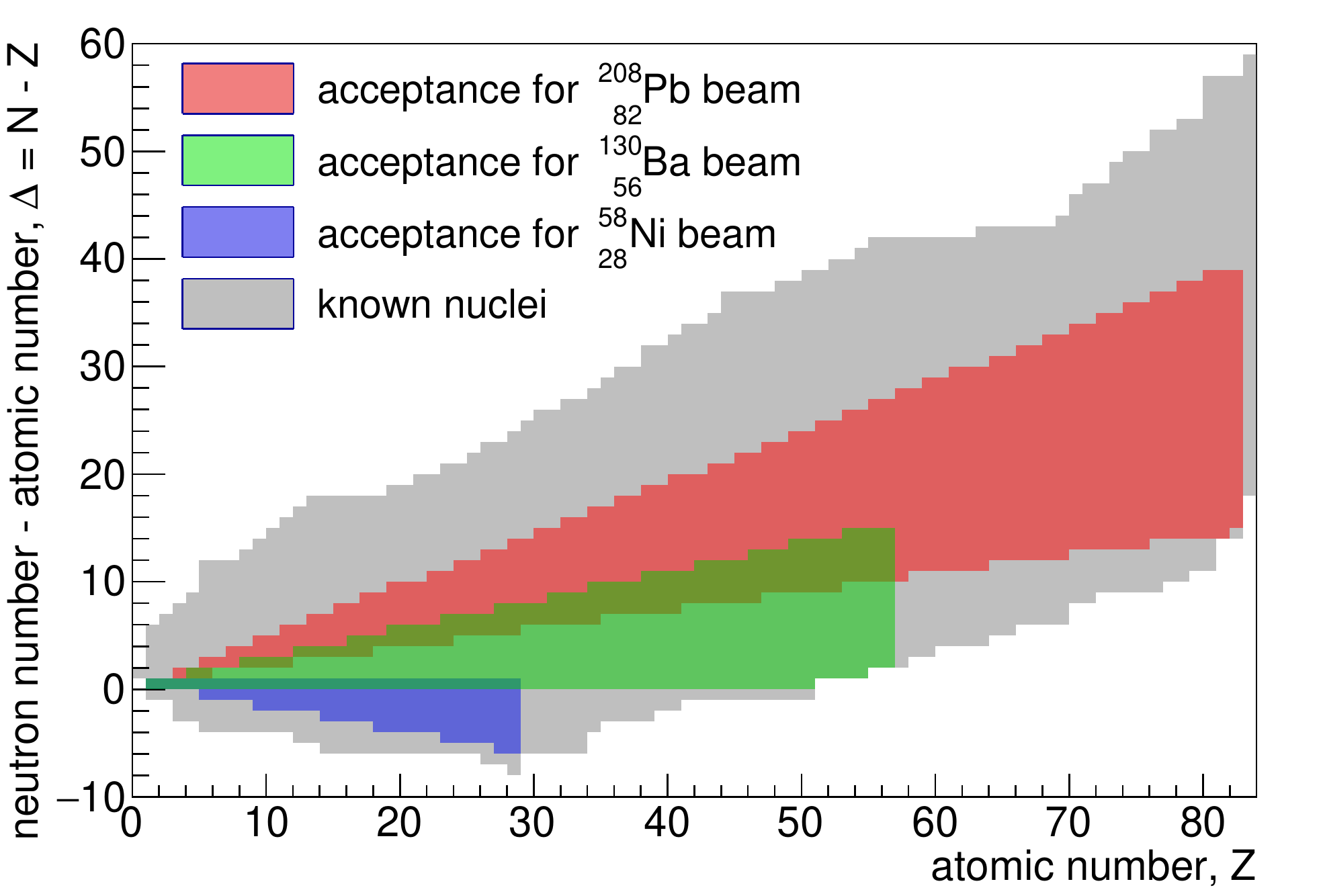}
  \caption{Approximate AFP acceptance regions for nuclear debris for different beam ions 
  (\ce{^{208}_{82}$Pb$}, \ce{^{130}_{56}$Ba$} and \ce{^{58}_{28}$Ni$}).}
  \label{fig:acceptance_xxx}
\end{figure}

If the aim of the measurement would be to study the abundances of different nuclei, yet another possibility to modify the acceptance region would be to change the beam ions.
Then, the $x=0$ positions would be defined by a different $A/Z$ value, and the acceptance would change accordingly.
Fig.~\ref{fig:acceptance_xxx} presents  the acceptance regions accessible when beams of 
\ce{^{208}_{82}$Pb$}, \ce{^{130}_{56}$Ba$} and \ce{^{58}_{28}$Ni$} are accelerated to energies equivalent to 6.5~TeV protons.
It was assumed that the optics is the same for all those scenarios.
It is a reasonable assumption, since the acceptance is determined mainly by the beam separation magnets, which are set the same for 
all optics.
Furthermore, the optics for $PbPb$ and $pp$ running at the LHC is very similar, so one should not expect problems for other beam ions either.
It can be observed that with the machine settings mentioned above the acceptance extends towards lower $A/Z$ values.
An extension toward larger $A/Z$ would be very difficult, since no appropriate stable nuclei exist.
The \ce{^{130}_{56}$Ba$} and \ce{^{58}_{28}$Ni$} nuclei were chosen as examples, since they are stable and together with \ce{^{208}_{82}$Pb$} they cover a very large region.

\section{Conclusions}
\label{sec:conclusions}

In the presented study it was shown that the existing forward proton detectors at the LHC provide an interesting possibility of detecting 
nuclear debris emerging from the collision of two heavy ions.
This possibility could be used for a measurement of abundances of various nuclei produced in heavy-ion collisions.

Another interesting goal would be a measurement of the centrality of the collision.
Different centralities would result in different signals generated by the created nuclear debris.
Such a measurement would be independent of and complementary to other commonly used methods (we leave this topic for a next study).

The standard methods of the centrality measurement can only be used to compare the centrality of two given events.
Such a procedure does not provide the absolute scale and leads to the concept of the \textit{centrality bins}.
Obviously, the present set-up of the AFP detectors alone would not help.
However, it is possible to consider similar detectors located in other places, in order to extend the joined acceptance to the whole spectrum of nuclei.
This, together with the zero degree calorimeters measuring the neutrons emitted in the extreme forward direction, could provide a direct measurement of 
the number of spectators and hence, allow the determination of the number of participants, see also~\cite{Tarafdar:2014oua}.
Contrary to the presently used methods, such a measurement could be model independent.

\section*{Acknowledgements}
We gratefully acknowledge discussions with S. Tarafdar and A. Milov on these topics.
This work was supported in part by Polish National Science Centre grant UMO-2012/05/B/ST2/02480.

\end{document}